# High Level Hardware/Software Embedded System Design with Redsharc

Sam Skalicky, Andrew G. Schmidt and Matthew French
Information Sciences Institute
University of Southern California
Arlington, VA USA
{skalicky,aschmidt,mfrench}@isi.edu

*Abstract*—As tools for designing multiple processor systems-on-chips (MPSoCs) continue to evolve to meet the demands of developers, there exist systematic gaps that must be bridged to provide a more cohesive hardware/software development environment. We present Redsharc to address these problems and enable: system generation, software/hardware compilation and synthesis, run-time control and execution of MPSoCs. The efforts presented in this paper extend our previous work to provide a rich API, build infrastructure, and runtime enabling developers to design a system of simultaneously executing kernels in software or hardware, that communicate seamlessly. In this work we take Redsharc further to support a broader class of applications across a larger number of devices requiring a more unified system development environment and build infrastructure. To accomplish this we leverage existing tools and extend Redsharc with build and control infrastructure to relieve the burden of system development allowing software programmers to focus their efforts on application and kernel development.

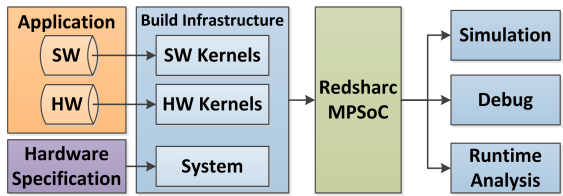

Figure 1: Redsharc implementation flow for MPSoC designs.

## I. Introduction

To the uninitiated, implementing applications with heterogeneous multiprocessor systems-on-chips (MPSoCs) can be quite a daunting task. Even for experienced developers designs are challenging and include sequential processor threads, parallel hardware accelerators, and system-wide integration resulting in a complex system-on-a-chip (SoC) infrastructure. In order to extract the best performance from a system, applications must most effectively utilize the available resources without requiring the developer to possess advanced degrees in computer architecture and years of experience with both software and hardware development.

These SoCs integrate the functionality previously implemented in separate chips into a single device, reducing component counts by using integrated processors, memory and high speed network interfaces, and a vast amount of on-chip programmable resources. Although tools exist that support system construction and assembly with the option to use pre-built hardware IP cores from vendors like Xilinx and Altera, they still require low level design knowledge and lack support for controlling and implementing the user's application.

Even as tools continue to evolve to better meet the demands of the developers, there still exist systematic gaps that must be bridged to provide a more cohesive hardware/software development environment. One approach taken recently by vendors is to provide more capable embedded processors, such as the ARM Cortex-A9 in Xilinx Zynq and Altera HPS, so developers can write software without needing to build an SoC and program the FPGA device. With the addition of High Level Synthesis (HLS) tools, a developer can quickly build a library of hardware IP cores to take advantage of the FPGA's vast heterogeneous, high-performance, resources.

Unfortunately, enabling a run-time system that can succinctly integrate multiple hardware and software compute components to best meet the needs of an application remains a challenging task. This problem is precisely the focus of this work. That is to say, *Can we automatically generate such systems from user specifications?* Given a system, compiling and synthesizing software and hardware kernels is relatively straightforward. But controlling a complex heterogeneous system is not so simple. What amount of direction must the designer provide in order to control the system — when should tasks start and data transfers occur?

We present Redsharc to address these problems and enable: system generation, software/hardware compilation/synthesis, and run-time control of MPSoCs. The Reconfigurable Data-Stream Hardware Software Architecture (Redsharc) has been previously introduced [1][2] as a solution to meet the performance needs of MPSoCs. The efforts presented in this paper extend our previous work to provide a rich API, build infrastructure, and runtime environment to allow developers to design a system of simultaneously executing kernels in software or hardware communicating across a seamless interface.

A high level view of the Redsharc development and implementation process is shown in Figure 1. The user provides the application design, as a set of kernel implementations in hardware and/or software, and a hardware specification for the composition of processors and hardware cores in the system. Redsharc then takes these, compiles, synthesizes, and generates the heterogeneous hardware/software MPSoC. The system produced is fully functional, requiring no further user input to setup, or configure the design. Users can take the executables and bitstreams and download them directly to the device to begin execution and evaluate their application.

The rest of this paper is organized as follows. Section II discusses the relevant related works to Redsharc and this paper. Section III presents design and implementation details for users of Redsharc, followed by an example application implementation in Section IV. Section V summarizes our contributions and describes our continuing efforts with Redsharc for the future.

## II. RELATED WORK

Redsharc encompasses system infrastructure in the form of on-chip networks, a simple set of APIs for kernel entry, reduces barrier to entry for hardware kernels by integrating existing high level synthesis (HLS) tools, and a set of system configuration APIs for the user to describe their system. Below we present relevant related works in these areas.

**System Design Frameworks.** Improving the design process for SoCs has been approached from a variety of different angles. Design flows have been presented based around a single processor core integrating various software threads and hardware co-processors such as LegUp [3] and hthreads [4] among others [5]. Extensions to hthreads include a vendor neutral soft processor design flow [6], an OpenCL approach [7], and extensions for partial reconfiguration in the HEMPS project [8]. Wachter *et al.* [9] presented a framework for packet-switched NoC based MPSoC systems but do not incorporate hardware cores into the system. Chung *et al.* [10] presented CoRAM as a hardware design support system providing hard IP for memory controllers, NoC, and a simplified control mechanism to manage the hardware. Their system does not support integrating processor cores or software threads.

**HLS Tools.** Large strides have been made towards reducing the time and effort to design hardware using HLS tools. A large variety of HLS tools now exist from commercial products such as Vivado HLS from Xilinx, Impulse C, and Synfora PICO Extreme HLS to open source tools developed from academic research initiatives, such as LegUp [3], Catapult C [11], and MyHDL [12] among others.

**System Design Incorporating HLS.** Rather than just easing hardware design with HLS, other works have also integrated system design to enable entire systems to be created to support the hardware implementations of software routines. Adler *et al.* [13] presented LEAP to configure not just the FPGA but also an application management and control interface by abstracting the interface between hardware and software. Altera's OpenCL HLS is a framework to implement a system on an FPGA with compute pipelines interfaced through PCIe to a main control routine running on a PC. In their work on the Catapult project Putnam *et al.* [14] also found that the support system for the compute logic could be standardized, creating their Shell/Role scheme. Compared to CoRAM, they customized the Shell specifically for their problem domain rather than pure general applicability.

While these efforts work to solve some of the problems in the design of MPSoCs, there are still gaps in the overall flow. Redsharc aims to fill these gaps by supporting both software kernels running on processor cores and independent hardware cores connected over a seamless, verified, scalable on-chip network with streaming and block-based communication.

## III. REDSHARC

The renewed efforts with Redsharc are aimed at providing both software and hardware designers a simplified development environment, shifting the focus from system design and integration to application and kernel development. This includes integrating: HLS to rapidly implement hardware accelerated kernels, automatic system control for user defined scheduling policies, and a build framework to generate the binaries needed to implement the system. Furthermore, Redsharc now supports a vendor-agnostic development environment, enabling migration across FPGA generations and vendor devices.

Redsharc is based on the Stream Virtual Machine API (SVM) [15], an intermediate language between high level stream languages and low level instruction sets of various architectures developed under DARPA's Polymorphous Computing Architectures (PCA) program. SVM has no preference to the computational model and only specifies how kernels communicate with each other. SVM is primarily based on a streaming model, but additionally includes supports for blocks, or random access chunks of data. Redsharc has been developed to implement the SVM API across the hardware/software boundary and provides a cohesive build and run-time environment to support FPGAs and MPSoCs.

### A. Fundamental Features

Redsharc addresses the challenges of achieving inter-core communication with support for different communication models. The goal is to support any configuration of heterogeneous hardware and software kernels to fit the needs of the application. Redsharc provides fast and scalable on-chip networks that implement the Redsharc API. In a Redsharc system, tasks are known as kernels and implemented as either software threads running on a processor, or hardware cores in the FPGA fabric. Regardless of whether a kernel runs on a processor or hardware core, or in which core it runs on in the system, all kernels communicate using Redsharc's abstract API supporting both streaming and transmission of blocks of data. These transmissions occur over the proven, validated, and configurable Redsharc on-chip networks.

The form of MPSoC systems that can be created using Redsharc is shown in Figure 2. Previous works supported multiple software kernels assigned to the same processor core. In addition, Redsharc now supports multiple kernels assigned to any type of core. For hardware cores, this means that the two kernels are physically implemented side-by-side in the same reconfigurable fabric. For software cores, this means that the two kernels are executing simultaneously on the same physical processor core — sharing compute time by context switching.

The stream switch network (SSN) and block switch network (BSN) allow data to be transmitted through different modes as needed by the application. The SSN is a runtime reconfigurable crossbar on-chip network designed to carry streams of data between cores. The BSN is a routable crossbar on-chip network that permits access to any blocks from any kernel. The BSN memories include a set of on-chip block-RAM (BRAM) memories and connections to off-chip memories such as SRAM or DDR. The type of memory allocated to a kernel (either BRAM or an allocation in an off-chip memory) enables the system to choose between memory speed and density to meet the needs of the application. More information regarding the SSN and BSN, including performance analyses, can be found in previous work [1][2].

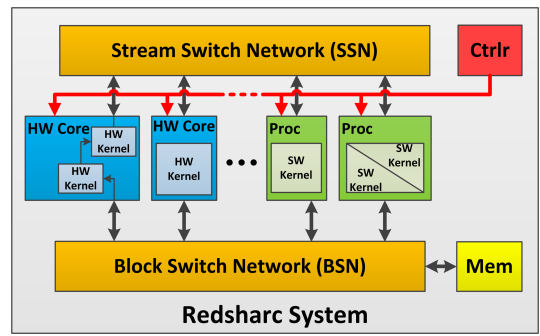

Figure 2: Example Redsharc MPSoC generated system.

*B. Developer Roles*

Generally, designing embedded systems requires a wide range of skills. Domain expertise is needed to understand the problem and craft a solution or algorithm. This algorithm will then need to be decomposed into kernels implemented on the type of core that meets its computational pattern and needs. At the system level, expertise is needed to determine the number of cores, types of cores —whether each is a processor or hardware core— and the policy to schedule kernels on the cores. This also includes device specific expertise to make sure the developed design properly takes advantage of the rich heterogeneous resources and I/O of the device. Figure 3 shows the differences in complexity and time/effort required for each of the different development duties.

Designing an entire heterogeneous hardware/software system from scratch normally entails allocating a majority of the time and effort to integration, testing, and verification at the system level, as shown in Figure 3a, and less on the kernel implementations and overall application design. In contrast, Redsharc reduces the need to have a strong skill set at the system level by providing proven and validated on-chip networks, communication interfaces and control necessary to manage execution. The addition of HLS enables the software kernels to be implemented in hardware reducing the complexity of the kernel implementations as shown in Figure 3b. Moreover, there are a multitude of HLS tools available that support languages including: C/C++ [3][16][17], Python [18], and Haskell [19] among many others [11][13] reducing the skills and time/effort needed to implement kernels.

*C. Kernel Development with Redsharc*

Application implementation begins by decomposing the application into kernels. These kernels can either be software threads or hardware logic. Then, leveraging the Redsharc API a developer can quickly assemble, generate, and test the system on the device. This approach allows for rapid development and testing along with providing vendor-agnostic implementations for ease of platform migration. Furthermore, as HLS tools continue to mature, the ability to rapidly integrate generated hardware kernels will further alleviate a software developers burden of hardware design.

In Redsharc the software kernel interface (SWKI) is implemented as a traditional software library. The SWKI provides an API for communication and data transfer, as shown in Table I, to other kernels via provided drivers to access the DMA controllers. A full description of the API calls is presented in [2]. Each type of processor may implement the SWKI in different ways. Software kernels are supported by a microkernel or small scale real-time operating system (RTOS) that interfaces between the on-chip networks, supports the management functions of the control kernel (starting, stopping, launching kernels), and enables context switching to support multiple simultaneously executing software kernels on the same processor. However, the RTOS is very thin providing direct access to driver routines enabling each kernel to run at full speed on the processor, only interrupting for context switching or as directed by the control kernel for management functions. The RTOS sets up and configures DMA, providing pointers for the software kernels to interact with directly.

Table I: Examples of Redsharc software kernel API calls.

| Function Name | Arguments | Description |
| --- | --- | --- |
| streamPush | | Pushes element e onto stream s |
| streamPop | element *e stream *s | Pops the top element from stream s and stores the value in e |
| streamPeek | | Reads the top element from stream s and stores the value in e |
| blockWrite | element *e int index block *b | Writes element e into block b at index i |
| blockRead | | Reads and element from block b at index i and stores the value in e |

The hardware kernel interface (HWKI) is a thin wrapper that connects hardware kernels to the SSN and BSN, implemented as a VHDL entity. The HWKI is composed of 3 sets of interfaces: control registers, blocks, and streams as shown in Figure 4. Control registers allow the control kernel to start, stop, and reset each core and enables the kernel to share status or debug information. The block interface connects directly to the BSN and provides a simple set of block RAM-like interfaces for the kernel to interact with. The stream interface connects directly to the SSN and provides standard FIFO interfaces. Specifically which block or stream each kernel is interacting with is handled separately by the control kernel and implemented by the BSN and SSN.

Redsharc supports the design of hardware kernel implementations using HLS by accepting the input software code, running the HLS tool to generate the core functionality and ensuring that the top level interface implements the Redsharc HWKI. This procedure enables developers with little hardware experience, or experienced hardware developers with little time, to design hardware implementations. Currently, Redsharc supports integration with Vivado HLS by augmenting the generated IP core with the HWKI. At present, a designer must select the appropriate directives to ensure BRAMs and FIFOs are the primary interface, but rather than requiring an AXI or bus-based top-level interface Redsharc uses Python scripts to integrate the necessary HWKI into the hardware design.

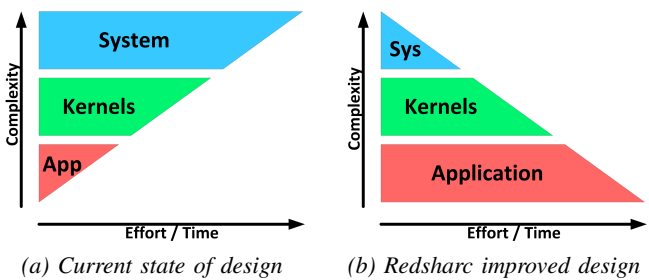

(a) Current state of design  (b) Redsharc improved design

Figure 3: Shifting development focus with Redsharc

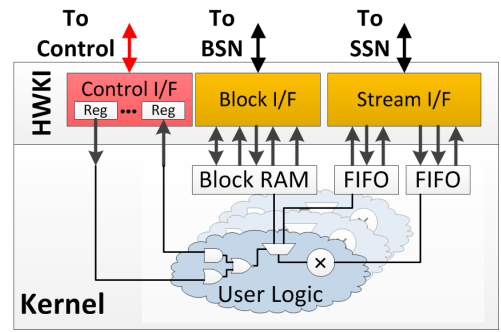

Figure 4: Hardware Kernel Interface API simplifies user designs with BRAM and FIFO-like interfaces for communication

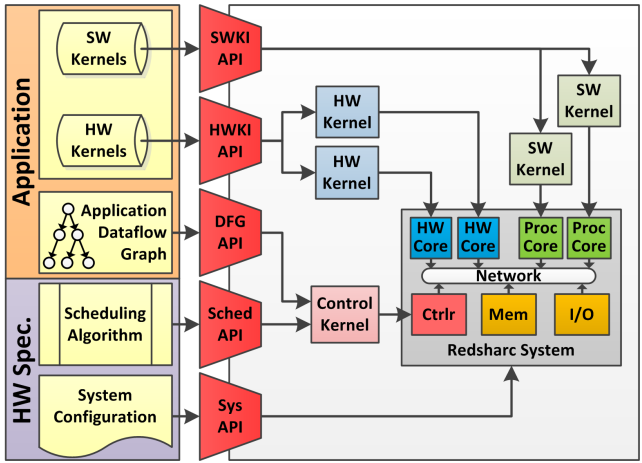

Figure 5: Overview of Redsharc API showing the various input data required, and how it is utilized to construct the system

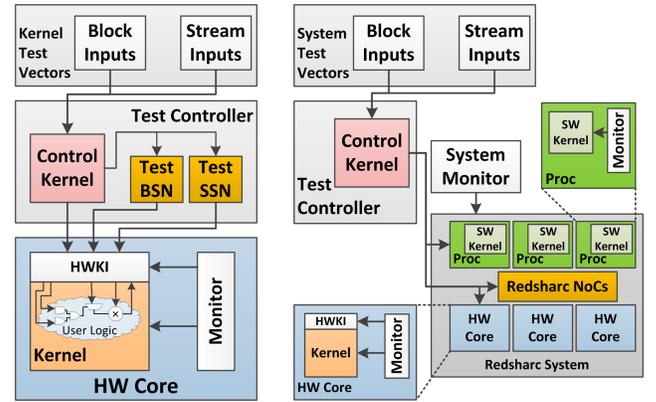

(a) HW Kernel Testbench  (b) Redsharc MPSoC Testbench

Figure 6: Template simulation testbenches for individual hardware kernel *(a)* and full/partial MPSoC system simulation *(b)*.

*D. System Development with Redsharc*

To design a system with Redsharc the designer supplies:
1) Kernel implementations
2) Dataflow graph (DFG) of the application
3) Scheduling policy to be applied
4) Configuration of the system

These inputs to Redsharc are depicted in yellow in Figure 5. The Redsharc APIs are shown in red. The control kernel operates using the provided DFG and scheduling policy, no other user intervention is required.

To reduce the skill set requirements Redsharc only requires developers to provide the set of dependencies between kernels, the configuration of the processing and hardware cores in the system, and the scheduling policy to implement. Redsharc provides a DFG API for specifying the dependencies between kernels in the form of a dataflow graph (DFG). In this graph each node represents a kernel and an edge is a data dependency between two kernels. During execution the DFG will be traversed by the control kernel to maintain correct operation of the system. The scheduling policy defines the order that kernels will be executed and to which core they will be assigned.

The system configuration includes the number of processor and hardware cores that will be available to execute tasks. Through the Redsharc System API, the user specifies the capabilities of each core and the number of stream and block interfaces that will be needed by each kernel. The actual implementation of the system will be generated from this configuration utilizing pre-designed processor blocks, hardware modules, and on-chip networks using a set of makefiles to interface directly with the vendor-supplied compilation tools.

When designing with Redsharc, the same integrated development environments (IDEs) and software development kits (SDKs) are used. For hardware design entry and validation, the same VHDL/Verilog IDE and simulators are used. Once the kernel designs are validated, the source code is provided to Redsharc and behind the scenes the same vendor supplied compilation tools (such as gcc or g++ for software, or xst, ngbuild, or bitgen for hardware) are called for compilation and synthesis of the kernel designs, described next.

*E. Build Infrastructure*

Part of Redsharc includes a build infrastructure to support rapid assembly, configuration, and testing of developed hardware kernels and full systems. The goal of the build infrastructure is to allow a developer to spend more time developing kernels, rather than creating test benches and simulation/synthesis project files.

The developer can leverage provided makefiles, simulation and synthesis scripts to rapidly simulate and synthesize a kernel for debugging and testing as shown in Figure 6a. With the Redsharc API and template simulation test bench the stream, block, and control transactions are managed for the developer. Input streams and blocks are provided as files to the simulation environment and output streams and blocks are checked against expected results for validation. The simulation environment currently supports Synopsys VCS. Multiple test vectors can be loaded into to simulation environment and can be used as regression tests while a kernel is under development.

The build infrastructure also supports the testing of multiple kernels assembled together as a subsystem or full system as shown in Figure 6b. This includes the use of pre-configured soft-core processors to emulate software kernels and pre-designed stream and block switch networks (SSN and BSN) for connectivity of the system. System simulation can be performed at various stages during the build process: pre-synthesis, post-synthesis, and post-PAR, including timing information. Full-system synthesis and implementation is also supported for both Xilinx and Altera systems, leveraging the Xilinx ISE[1] and Altera Quartus II tool chains.

Given the kernel implementations, dependencies, system configuration, and scheduling policy Redsharc composes a control kernel to manage communication and execution at runtime. Previously, control kernels were only implemented in software for simplicity. However, in this work we removed this restriction and introduced simplifications to the API to ease the user effort and enable a standardized hardware kernel to control the system.

*F. System Runtime Operation*

After a system has been designed and implemented, the next task is to get it up and running. After the initial bitstreams and processor executables have been downloaded, the control kernel begins setting up BSN/SSN network connections, scheduling and launching worker kernels to execute parts of the application. Kernels are assigned to the processor or hardware core as specified by the scheduling policy and following the dependencies in the DFG to ensure correct execution. Before a kernel is started, its block and stream interfaces are configured in the BSN and SSN appropriately.

---
[1]Development is underway to also support Vivado

After execution has begun no more user interaction is required. The control kernel frees block and stream resources when both the kernel putting in data and the kernel reading out the data have finished. Then these resources are used to support communication in other kernels. The control kernel can be monitored by the user and, signal when final data has been produced or when all kernels have finished executing.

After a software kernel finishes executing, the RTOS running on the local processor core frees up any private resources allocated, allowing other kernels to use them. However launching another hardware kernel is not so simple. To achieve the same functionality we leverage partial reconfiguration to reconfigure the FPGA fabric for the incoming hardware kernel. Just as with the processor cores that have a hardware limited number of DMA controllers, hardware cores have only a fixed number of physical block and stream ports that connect to the BSN and SSN. The HWKI supports more block and FIFO interfaces by buffering and interleaving data on a single physical channel. The specific configuration of the HWKI is generated by Redsharc automatically during implementation and synthesis based on the system specification.

Once a new system design has been completed the next question is: "Is the performance of the system what I expect?" In our previous work, we have developed an extensible performance monitoring infrastructure [20]. By leveraging this framework, Redsharc provides two types of system generation: Analysis (for performance monitoring), and Release (without the performance monitoring framework). Additionally, debug functionality of the system can be had through a system configuration setting to direct Redsharc to include debug capabilities in the control kernel. Through the control kernel, the user can "pause" execution, read/modify current data in blocks and streams, and other debug functions as necessary.

## IV. EXAMPLE APPLICATION IN REDSHARC

To demonstrate the simplicity and ease of use of Redsharc, we present an example face recognition application and show the steps required for implementation. In this section we present an overview of the face recognition algorithm and how it was implemented with sample kernel implementations, kernel setup, and configuration in the DFG.

Facial recognition is often used in consumer products like Google Picasa, Microsoft Live Gallery, or Facebook and in law enforcement or military intelligence to identify a person of interest. To process this massive amount of data, reductions in dimensionality are necessary to effectively analyze as many images as possible. One way to achieve this is to extract the most important features from the image, producing eigenfaces as introduced by Turk and Pentland [21]. In their approach principal component analysis (PCA) is used to produce the eigenvalues for the image, making up the eigenface. We use singular value decomposition (SVD) to perform PCA. The determination if a sample face matches a subject in the reference database is calculated by computing the feature vector for that sample face. Then the root-mean-square (RMS) differences between the sample face's feature vector and the feature vectors for the reference subjects is computed. The closest matching subject is the one with the lowest RMS difference. Figure 7 shows the DFG for this application.

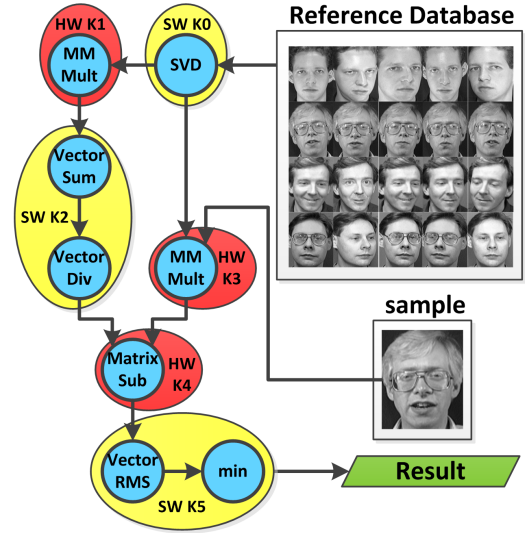

Figure 7: Face Recognition DFG partitioned into software and hardware kernels.

Implementing this face recognition application using Redsharc is a step-by-step progression of migrating the existing implementation to Redsharc primitives which simplify implementation. First we began with an initial sequential C code application. Then, the code for each kernel was segmented into separate functions and the shared data variables were moved to a global scope. These global variables were then reimplemented as Redsharc blocks and streams. Converting the functions into software kernels using the SWKI was a simple matter of migrating access to the global blocks and streams into local blocks and streams passed as an argument into each kernel as shown in Listing 1. At this point each kernel is independent of any other kernel, simply reading and modifying the given data structures. The DFG was implemented using the API by specifying which blocks and streams are produced by one kernel and consumed by another, as shown in Listing 2. The Redsharc implementation of this application produced

```
1  void swk4(struct taskData *data) {
2      // get references to the data structures
3      redsharc_block *pc1 = data->blocks[0];
4      redsharc_stream *mean = data->streams[0];
5      redsharc_stream *diff = data->streams[1];
6      int i,j;              // do work
7      for(i=0; i<m; i++) {
8          for(j=0; j<n; j++) {
9              double tmp0,tmp1,tmp2;
10             blockRead(&tmp0,j,pc1);     // from hwk3
11             streamPop(&tmp1,mean);      // from swk2
12             tmp2 = tmp0 − tmp1;
13             streamPush(&tmp2,diff);     // to swk5
14         }
15     }
16     notify_kernelFinished(data->handle);
17 }
```

Listing 1: Software Kernel 4 Implementation

```
1  //setup kernel 4 with two inputs and one output
2  initKernel(4, HW4, 2, 1, dfg);
3  //setup first input as a stream from kernel 2
4  addStreamDependency(4, 0, 2, 0, dfg);
5  //setup second input as a block from kernel 3
6  addBlockDependency(4, 1, 3, 0, dfg);
7  //setup first output as a stream
8  addOutputStream(4, 0, DOUBLE, N*totalImages, dfg);
```

Listing 2: DFG API for Configuring Kernel 4

exactly the same results for a variety of sample images when compared to a given reference database as the initial single threaded C code implementation.

The same user provided DFG and kernel implementations can be run on any system with any number of processor cores with a simple change to the system configuration. Interesting future work would be to implement some of the kernels in hardware either manually or using HLS and compare the performance of various system configurations for a variety of applications. In addition, we are also working towards providing support to various vendor architectures from standard C pthreads for initial testing, Xilinx Zynq/PPC, Altera Nios/HPS, and ARM soft-cores on Achronix FPGAs.

## V. Conclusions and Ongoing Work

This paper presents an improved Reconfigurable Data-Stream Hardware Software Architecture (Redsharc) infrastructure greatly expanded to support software development and reduce system design efforts. Redsharc leverages our stream and block switch networks with a goal to reduce hardware kernel development complexities, allowing developers to spend more time on kernel development. Taking Redsharc further to support a broader class of applications across more devices required a more cohesive system development environment and build infrastructure. The additions to scheduling and control of the system relieve much of the system development burden, allowing the software developers to focus their efforts on application and kernel development.

Although Redsharc has been improved significantly, we have not yet achieved our goal of an end-to-end framework for vendor agnostic hardware/software MPSoC development. In this effort, most of our improvements have been software and hardware infrastructure. Since a benchmark suite of kernels running in software and hardware primarily evaluates the HLS tools (or the device), a more useful analysis would be across different FPGA platforms. Other work of interest includes better support for Partial Reconfiguration to further improve hardware resource utilization. Towards this goal we are working to incorporate Torc's MicroBitstream generation [22] and better integrating the HLS flow with a PR flow. These works are all aimed to be released as Open Source; at the time of writing a date has not yet been set.

## References

[1] A. Schmidt, W. Kritikos, R. Sass, E. Anderson, and M. French, "Merging Programming Models and On-chip Networks to Meet the Programmable and Performance Needs of Multi-core Systems on a Programmable Chip," *International Conference on Reconfigurable Computing and FPGAs*, Dec. 2010.

[2] W. Kritikos, A. Schmidt, R. Sass, E. Anderson, and M. French, "Redsharc: A Programming Model and On-Chip Network for Multi-Core Systems on a Programmable Chip," *International Journal of Reconfigurable Computing*, 2012.

[3] A. Canis, J. Choi, M. Aldham, V. Zhang, A. Kammoona, T. Czajkowski, S. D. Brown, and J. H. Anderson, "LegUp: An Open-source High-level Synthesis Tool for FPGA-based Processor/Accelerator Systems," *ACM Transactions on Embedded Computing Systems*, vol. 13, no. 2, Sep. 2013.

[4] D. Andrews, D. Niehaus, R. Jidin, M. Finley, W. Peck, M. Frisbie, J. Ortiz, E. Komp, and P. Ashenden, "Programming Models for Hybrid FPGA-CPU Computational Components: A Missing Link," *IEEE Micro*, vol. 24, no. 4, July 2004.

[5] S. S. Bhattacharyya, G. Brebner, J. W. Janneck, J. Eker, C. von Platen, M. Mattavelli, and M. Raulet, "OpenDF: A Dataflow Toolset for Reconfigurable Hardware and Multicore Systems," *SIGARCH Computer Architecture News*, vol. 36, no. 5, June 2009.

[6] E. Cartwright, A. Fahkari, S. Ma, C. Smith, M. Huang, D. Andrews, and J. Agron, "Automating the Design of mLUT MPSoPC FPGAs in the Cloud," *International Conference on Field Programmable Logic and Applications*, Aug. 2012.

[7] S. Ma, M. Huang, and D. Andrews, "Developing Application-specific Multiprocessor Platforms on FPGAs," *International Conference on Reconfigurable Computing and FPGAs*, Dec. 2012.

[8] E. Cartwright, A. Sadeqian, S. Ma, and D. Andrews, "Achieving Portability and Efficiency over Chip Heterogeneous Multiprocessor Systems," *International Conference on Field Programmable Logic and Applications*, Sept. 2014.

[9] E. Wachter, C. Lucas, E. Carara, and F. Moraes, "An Open-Source Framework for Heterogeneous MPSoC Generation," *Southern Conference on Programmable Logic*, Mar. 2012.

[10] E. S. Chung, J. C. Hoe, and K. Mai, "CoRAM: An In-fabric Memory Architecture for FPGA-based Computing," *ACM/SIGDA International Symposium on Field Programmable Gate Arrays*, Feb. 2011.

[11] C. Economakos and G. Economakos, "FPGA Implementation of PLC Programs Using Automated High-Level Synthesis Tools," *IEEE International Symposium on Industrial Electronics*, June 2008.

[12] J. I. Villar, J. Juan, M. Bellido, J. Viejo, D. Guerrero, and J. Decaluwe, "Python as a Hardware Description Language: A Case Study," *Southern Conference on Programmable Logic*, Apr. 2011.

[13] M. Adler, K. E. Fleming, A. Parashar, M. Pellauer, and J. Emer, "Leap Scratchpads: Automatic Memory and Cache Management for Reconfigurable Logic," *ACM/SIGDA International Symposium on Field Programmable Gate Arrays*, Feb. 2011.

[14] A. Putnam, A. Caulfield, E. Chung, D. Chiou, K. Constantinides, J. Demme, H. Esmaeilzadeh, J. Fowers, G.P. Gopal *et al.*, "A Reconfigurable Fabric for Accelerating Large-Scale Datacenter Services," *International Symposium on Computer Architecture (ISCA)*, June 2014.

[15] P. Mattison and W. Thies, "Streaming virtual machine specification, version 1.2, technical report," January 2007.

[16] K. Denolf, S. Neuendorffer, and K. Vissers, "Using C-To-Gates To Program Streaming Image Processing Kernels Efficiently on FPGAs," *International Conference on Field Programmable Logic and Applications*, Aug. 2009.

[17] J. Xu, N. Subramanian, A. Alessio, and S. Hauck, "Impulse C vs. VHDL for Accelerating Tomographic Reconstruction," *IEEE International Sympoisum on Field-Programmable Custom Computing Machines*, May 2010.

[18] G. Inggs, D. Thomas, and S. Winberg, "Exploring the Latency-Resource Trade-off for the Discrete Fourier Transform on the FPGA," *International Conference on Field Programmable Logic and Applications*, Aug. 2012.

[19] S. M. Loo, B. E. Wells, N. Freije, and J. Kulick, "Handel-C for Rapid Prototyping of VLSI Coprocessors for Real Time Systems," *Southeastern Symposium on System Theory*, Mar. 2002.

[20] A. Schmidt, N. Steiner, M. French, and R. Sass, "HwPMI: An Extensible Performance Monitoring Infrastructure for Improving Hardware Design and Productivity on FPGAs," *International Journal of Reconfigurable Computing*, 2012.

[21] M. Turk and A. Pentland, "Eigenfaces for Recognition," *Journal of Cognitive Neuroscience*, vol. 3, no. 1, Jan. 1991.

[22] R. K. Soni, N. Steiner, and M. French, "Open-Source Bitstream Generation," *IEEE International Symposium on Field-Programmable Custom Computing Machines*, Apr. 2013.